\documentclass{aastex}
\usepackage{emulateapj5}

\def\gtrsim{\mathrel{\hbox{\rlap{\hbox{\lower4pt\hbox{$\sim$}}}\hbox{$>$}}}}
\def\lesssim{\mathrel{\hbox{\rlap{\hbox{\lower4pt\hbox{$\sim$}}}\hbox{$<$}}}}

\shortauthors{Lister}
\shorttitle{Relativistic Beaming and Flux Variability in AGNs}
\begin{document}
\slugcomment{Accepted for publication in the Astrophysical Journal}
\title{Relativistic Beaming and Flux Variability in \\
Active Galactic Nuclei}
\author{Matthew L. Lister}

\affil{National Radio Astronomy Observatory,\\
 520 Edgemont Road, Charlottesville, VA 22903-2454}
\email{mlister@nrao.edu}

\begin{abstract}
We discuss the impact of special relativistic effects on the observed
light curves and variability duty cycles of radio-loud active galactic
nuclei (AGNs). We model the properties of AGN light curves at radio
wavelengths using a simulated shot noise process in which the
occurrence of major flaring events in a relativistic jet is governed
by Poisson statistics. We show that flaring sources whose radiation is
highly beamed toward us are able to reach very high flux levels, but
will in fact spend most of their time in relatively low flaring
states. This is primarily due to relativistic Doppler contraction of
flaring time scales in the observer frame. The fact that highly beamed
AGNs are not observed to return to a steady-state quiescent level
between flares implies that their weakly beamed counterparts should
have highly stable flux densities that result from a superposition of
many long-term, low-amplitude flares. The ``apparent'' quiescent flux
levels of these weakly beamed AGNs (identified in many unified models
as radio galaxies) will be significantly higher than their ''true''
quiescent (i.e., non-flaring) flux levels.  We have also performed
Monte Carlo simulations to examine how relativistic beaming and source
variability bias the selection statistics of flat-spectrum AGN
samples. We find that in the case of the Caltech-Jodrell Flat-spectrum
survey (CJ-F), the predicted orientation bias towards jets seen end-on
is weakened if the parent population is variable, since the highly
beamed sources have a stronger tendency to be found in low flaring
states. This effect is small, however, due to the fact that highly
beamed sources are relatively rare, and in most cases their flux
densities will be boosted sufficiently above the survey limit such
that they will be selected regardless of their flaring level. We find
that for larger flat-spectrum AGN surveys with fainter flux density
cutoffs, variability should not be an appreciable source of selection
bias.

\end{abstract}

\keywords{relativity --- galaxies : active --- quasars : general ---
radio continuum : galaxies}
  
\section{Introduction}

One of the hallmarks of flat-spectrum radio-loud active galactic
nuclei (AGNs) is their tendency to display  large flux density
variations over a wide range of wavelengths. Studies of complete AGN
samples (e.g., \citealt{AAH92, LV99, LTP01}) have shown that the
degree of variability at radio wavelengths is well-correlated with the
prominence of a bright, flat-spectrum core component that is thought
to harbor the supermassive black hole and accretion disk that power
the AGN.  Parsec-scale images made with VLBI techniques show that this
core is usually located at the base of a highly collimated,
relativistic outflow. 

The extremely high speeds of AGN jets are responsible for numerous
biases in the observed properties of samples selected on the basis
of core flux density, due to special relativistic effects (e.g.,
\citealt{SR79}). For example, the distribution of source
orientations in such samples will be heavily weighted towards jets
seen nearly end-on, due to relativistic beaming of radiation in the
direction of motion. These highly beamed sources (also known as
blazars) will also tend to have more variable light curves than
typical radio-loud AGNs, since any intrinsic flux variations
associated with the relativistic outflow will have their timescales
shortened and amplitudes boosted in the observer's frame.

Previous theoretical studies of observational biases associated with
relativistic beaming in AGNs (e.g., \citealt{VC94, LM97}) have
considered only steady-state fluxes, and have not taken into account
possible additional biases due to variability. In most
flux-limited samples of astronomical objects, source variability is
not usually a major source of bias, since its effects tend to be
statistically averaged out in the sampling process. The situation is
different, however, for samples of objects whose flux densities are
relativistically beamed, such as flat-spectrum AGNs and gamma-ray
bursts. As we will show in this paper, the observed variability duty
cycles of these objects (i.e., the fraction of time spent above a
particular flux level) can be strongly affected by beaming, and should
therefore be considered as a possible source of bias in flux-limited
samples. Since these types of samples are often used to infer general
properties of the parent population, it is important to ascertain the
strength of this effect.

In the first part of this paper we develop a simple shot-noise
variability model to examine the effects of relativistic beaming on
the radio light curves and duty cycles of AGNs. We describe the
parameters of our model in \S~2, and compare our simulations to
observed AGN light curves in \S~3. In \S~4 we show how highly beamed
AGNs are able to reach very high flux levels, but will in fact be
observed to spend most of their time in low flaring states. In the
remaining portion of the paper we use Monte Carlo simulations to model
the beaming and variability properties of the blazar parent
population, and determine the degree to which core-selected AGN
samples are additionally biased by source variability. In \S~5 we show
that despite the dramatic range of {\it observed} variability levels
in blazars, their overall parent population likely consists of weakly
beamed sources with stable flux levels that arise from the
superposition of many long-term flares. We find that as a result, the
``quiescent'' flux levels of these sources (identified in most unified
models as radio galaxies) are significantly higher than their ``true''
quiescent (i.e., non-flaring) levels.  We further demonstrate how the
predicted orientation bias in medium-sized, flat-spectrum AGN samples
is slightly weakened in the case of a flaring source population. We
summarize our main findings in \S~6.

Throughout this paper we use a Freidmann cosmology with zero
cosmological constant, $q_o = 0.1$, and a Hubble constant $h = 0.65$
measured in units of $100 \rm \; km \; s^{-1}\; Mpc^{-1}$. Over the
redshift range of our simulations, this choice of cosmology gives
luminosity distances that are within $\sim 10$\% of those calculated
using more contemporary $\Omega_m = 0.28$, $\Omega_\Lambda = 0.72$
models (e.g., \citealt*{P99}).

\section{Variability Model}

The study of variability in blazars at radio wavelengths is hindered
by complications arising from relativistic beaming. The true jet
speeds and orientations of individual objects are in most cases poorly
known, which makes it difficult to disentangle possible beaming
effects from their observed light curves. As a result, there are
relatively few observational constraints on the intrinsic amplitudes
and timescales of individual flares that are responsible for the
majority of flux density variations in radio-loud AGNs.

These considerations have led us to develop a relatively simple
stochastic variability model for this study that contains a minimal
number of free parameters, yet still reproduces the major variability
characteristics of AGNs at radio wavelengths. In this
section we describe the parameters of this model, which we will use to
create simulated light curves for AGNs having various jet speeds and
orientations.

\subsection{Rate of flare occurrence\label{flarerates}}
 
Long-term flux monitoring studies at cm- and mm-wavelengths at the
U. Michigan \citep{HAA92} and Mets\"ahovi \citep{VTUN92} observatories
have shown that the flux density variations of AGNs tend to be
stochastic (i.e., there are very few instances of periodicity). Rather
than being completely random, however, AGN light curves display a
flare-like behavior whose spectral properties are consistent with a
shot-noise process \citep{CD85, HAA92, HB92}.  One way to reproduce
shot noise is through a superposition of a series of identical
impulses, occurring at intervals dictated by Poisson statistics.  In a
Poisson process, the overall rate of events is statistically constant,
yet the starting times of individual events are independent of all
previous ones.  The time intervals between events follow an
exponential distribution. We will use such a process in our AGN
variability model, by assuming a constant flare rate $\rho$, such that
the probability that a second flare will occur within a time interval
$\tau$ after the first one is $p(\tau) = 1 -
\exp{(-\rho \tau)}$.

Long-term monitoring of bright AGNs in the mm-wave regime at the
Mets\"ahovi radio observatory has shown that on average, most sources
experience one major flare per year (E. Valtaoja, private
communication).  After correcting for the mean redshift of the
Mets\"ahovi monitoring sample ($\bar z \simeq 1$; \citealt{LV99}),
this translates into a source rest frame flare rate of $\rho = (1+\bar z)
\rho_{obs} = 2 \;{\rm y}^{-1}$. For simplicity we will assume this
rate for all our simulated light curves.

\subsection{Flare intensity profile\label{flareprofile}}

VLBI monitoring of AGN jets on parsec scales has shown that flaring
events are often accompanied by the emergence of a bright component
from the base of the jet (e.g., \citealt{JMMA01}). For our purposes we
will assume that the source of an AGN flare is a luminous blob moving
down a relativistic jet at an angle $\theta$ to the line of sight with
velocity $\beta$ in units of the speed of light. The Doppler factor of
the blob is $\delta = (\Gamma -
\sqrt{\Gamma^2-1}\;\cos{\theta})^{-1}$, and the Lorentz factor is
$\Gamma = (1-\beta^2)^{-1/2}$. Standard relativistic transformations
(e.g., \citealt{BK79}) can be used to convert quantities in the
blob frame to the observer frame. In particular, the intrinsic time
scale of the flare is multiplied by a factor of $(1+z)/\delta$, while
its flux density $S_\nu$ is boosted by a factor $\delta^{3}$, where we
have assumed a spectral index $\alpha = 0$.  For convenience, we will
parameterize both the amplitude of the flare and the total flux
density of the source in units of the observed quiescent (non-flaring)
flux density $S_q$. We will assume that the latter has spectral index
of zero and arises in a continuous jet having the same Doppler factor as
the blob. Its flux density is therefore boosted by a factor 
$\delta^{2}$ (the flare emission has a larger Doppler boost due to its
finite lifetime --- see, e.g., Appendix B of \citealt*{UP95}). The observed source flux density at time $t$ since the
start of the flare is then the sum of the quiescent and flaring
components: $ S(t) = (\delta^2 S_q +
\delta^3 F(t) S_q ) / (\delta^2 S_q) = 1 + \delta F(t)$, where $F(t)$
represents a flare profile that ranges from 0 to $P$ times the
quiescent flux density in the source frame.

\begin{figure*}
\plotone{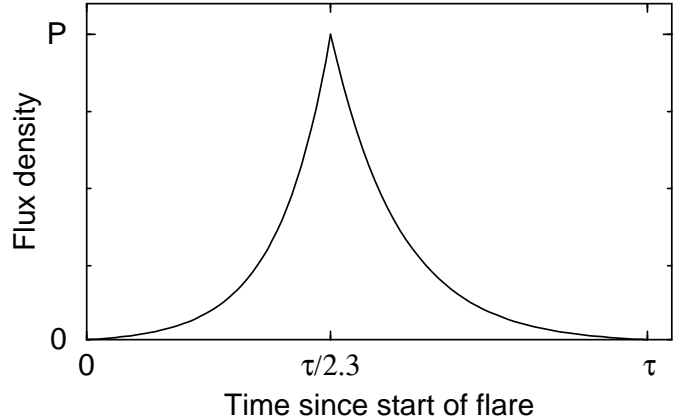}
\caption{\label{flareshape} Canonical flare intensity profile used in
simulated light curves, of the form $\bf\log{S} \propto t$, with
$t_{\rm decay} = 1.3 \; t_{\rm rise}$. The flare intensity is plotted in units of
the quiescent (non-flaring) flux density. }
\end{figure*}

\cite{TV94} have found that most major AGN flares in the mm-wave
regime have profiles of the form $\log{S} \propto t$ for the rise and
decay portions of the flare event. Fitting to more recent data by
\cite{VLT99} has shown that the decay time is typically a factor of
$\sim 1.3$ times longer than the rise time. In Figure~\ref{flareshape}
we display this profile, which has the form:

\begin{equation}
F(t) = \left\{ \begin{array}{ll} 
 {P\over(k-1)}\left[k^{2.3t/\tau}-1\right], & \mbox{$t \in [0,\tau/2.3]$}\\
 {P\over(k-1)}\left[k^{{2.3\over1.3}(1-t/\tau)}-1\right], & \mbox{$t
\in [\tau/2.3, \tau]$} \\
 0, & \mbox{$t > \tau$},
\end{array}
\right.
\end{equation}

where $\tau$ is the time scale of
the flare and $k$ is a constant.  One aspect of this profile is that
for low values of $k$, there are large exponential ``wings'' where the
flux density remains quite low compared to the peak. In order to make
$\tau$ more representative of the true time scale of the flare, we
minimize these wings by arbitrarily setting $k = 100$.

\subsection{\label{amptime}Flare timescales}

It is difficult to obtain good observational constraints on the range
of intrinsic timescales and amplitudes of AGN flares at radio
wavelengths, since isolated flares are exceedingly rare, and the
beaming factors are usually not well known. The majority of blazar
light curves are composed of many different flares having a variety of
timescales, so that often what appears to be a large isolated flare is
actually a superposition of several smaller events.

Given these uncertainties, we let the flare lengths in our model
be uniformly distributed over a range [$\tau_{min}$, $\tau_{max}$] in
the rest frame of the jet. Since our estimate of the flare rate ($2
\;\rm y^{-1}$) is based on bi-weekly monitoring data, and does not
account for any flares with observed timescales $\lesssim 1$ month we
set $\tau_{min} = 0.083$ y.

We derive a lower limit on $\tau_{max}$ by noting that there are no
sources in either of the two major AGN monitoring campaigns (U. of
Michigan or Mets\"ahovi) that show a tendency to return to a quiescent
baseline level in between flaring events. This implies that flares are
of sufficient duration in the source frame so that AGNs are rarely
observed in a non-flaring state. In our shot noise model, the
probability of this occurring is approximately
\begin{equation}
\label{fractime}
p \simeq e^{-\rho \bar\tau/\delta},
\end{equation}
where $\bar\tau \; = (\tau_{max}-\tau_{min})/2$ is the mean flare
time scale in the source frame. We can estimate an upper limit on
$\delta$ for the AGN parent population based on measurements of
superluminal motion in samples of highly beamed AGNs (i.e.,
blazars). 
\cite{JMM01} find component velocities in the jets of gamma-ray loud
blazars that range up to $\beta_{app}
\simeq 30 \; h^{-1}$. The relevant constraints from superluminal
motion equations ($\beta_{app} <
\Gamma$ and $\delta \lesssim 2\Gamma$; see \citealt{UP95}) imply a maximum
Doppler factor of $\sim 90$, assuming $h = 0.65$.  By choosing a
reasonably low probability for being in a non-flaring state ($p =
0.001$) and substituting $\delta = 90$, $\rho = 2\;\rm y^{-1}$ and
$\tau_{min} = 0.083
\rm \; y$ in Eq.~\ref{fractime}, we obtain $\tau_{max} \simeq 620$
y. Although this value is only a lower limit, we will show in \S~4 that
higher values of $\tau_{max}$ will simply increase the mean
steady-state flux level of the overall parent population, and will not
affect our main conclusions.

\subsection{Flare Amplitudes}
Given the observational uncertainties regarding the intrinsic flare
amplitudes of AGNs, we will make the simple assumption that all flare
amplitudes are the same fraction $P$ of a source's rest frame
quiescent flux level. Although this is likely a gross
oversimplification in the case of real AGNs, our study is not
concerned with low-amplitude flares, since they are not likely to
affect the overall selection statistics of flux limited
samples. Furthermore, we will show in \S~3 that a broad intrinsic flare
amplitude distribution is not needed to reproduce the basic properties
of AGN light curves, since the effects of flare superposition and
beaming can conspire to create a large range of observed flare
amplitudes.

We  constrain our flare amplitude parameter $P$ by examining the
range of observed flux density in the long term ($\sim 20$ yr) light
curves in the U. Michigan AGN sample. Of the 69 AGN light curves
analyzed by \cite{AAH99}, the highest absolute peak-to-minimum flux
ratio of any source at 4.8 GHz was $\sim 30$ for the BL Lac object
0235+164. Through repeated simulations of twenty-year light curves
with various values of $P$, we find that for $P = 0.05$, the
probability of finding a peak-to-minimum ratio exceeding 30 in a
highly beamed source ($\delta = 90$) is $\lesssim 0.001$. We therefore
adopt $P = 0.05$ in the simulations that follow.

\section{\label{comparisons}Comparisons to AGN light curves}

\begin{figure*}
\plotone{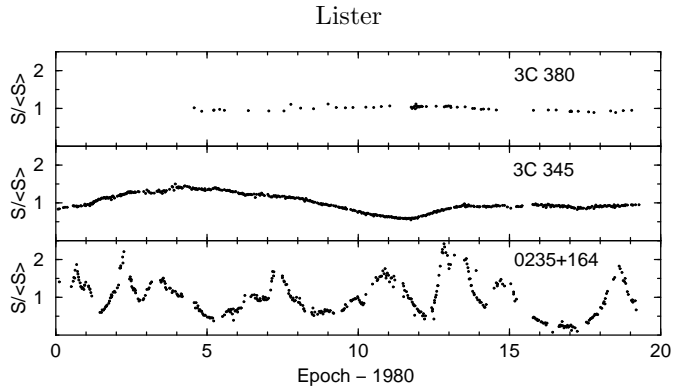}
\caption{\label{data_lc} Long-term light curves of three AGNs
at 4.8 GHz, from monitoring observations at the University of Michigan
Radio Observatory. The fluxes have been divided by the mean in each
case.}
\end{figure*}

Having established the general parameters of our model, we
are now able to compare our simulated light curves to those of actual
AGNs. In Figure~\ref{data_lc} we show the 20 year light curves of
three AGNs at 4.8 GHz from the University of Michigan AGN monitoring
program. In each panel the flux densities have been divided by the
mean for comparison purposes. The top, middle and bottom panels show
the light curves of the radio galaxy 3C~380, the quasar 3C~345, and
the BL Lac object 0235+164, respectively.

\begin{figure*}
\plotone{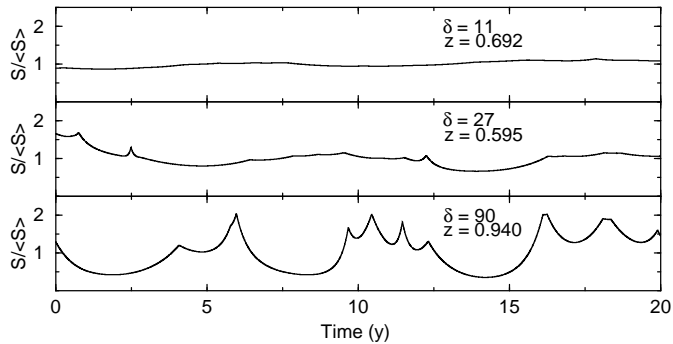}
\caption{\label{sim_lc} Simulated radio light curves of three AGNs
with redshifts and Doppler factors corresponding to the sources in
Figure~\ref{data_lc}. }
\end{figure*}

In order to make comparisons with our simulated light curves,
estimates of the Doppler factors of these sources are needed.  It is
possible to use the measured apparent speed in the inner jet of 3C~380
($\beta_{app} = 10.6$ ; \citealt*{PW98}) to obtain a crude estimate of
its Doppler factor if we assume that the jet is viewed at an angle
which maximizes its apparent velocity. This
gives $\delta = \beta_{app}/\beta \simeq \beta_{app} \simeq 11$.  The Doppler factor of 3C~345
has been well-constrained by \cite{WCR94}, who modeled the
three-dimensional motion of two components in its inner jet. They were
able to constrain the viewing angle to $\theta \simeq 2\arcdeg$ and
the Lorentz factor of the jet to be $\Gamma \gtrsim 20$, which implies
$\delta \gtrsim 27$. The BL Lac object 0235+164 is the one of the most
variable blazars in the U. Michigan monitoring sample, and is
therefore likely to have a Doppler factor at the high end of the
possible range ($\delta \sim 90$; see \S~\ref{amptime}). This source
has also been constrained by \cite{FKW99} to have $\delta > 80$ based
on equipartition arguments and the ratio of observed inverse-Compton
to synchrotron flux densities.

In Figure~\ref{sim_lc} we show three simulated light curves with
Doppler factors and redshifts corresponding to the three AGNs in
Figure~\ref{data_lc}. These light curves were calculated after
allowing the sources to achieve a steady-state flaring level. The
simulated flux densities have been divided by the mean in each
case. We note that the source redshift will affect the timescales but
not the amplitudes of our simulated light curves, since the flux
densities are measured in units of the observed quiescent level.  Our
simulated curves reproduce the general variability characteristics of
all three sources fairly well, even with our simplistic assumption of
equal flare amplitudes. 

\begin{figure*}
\plotone{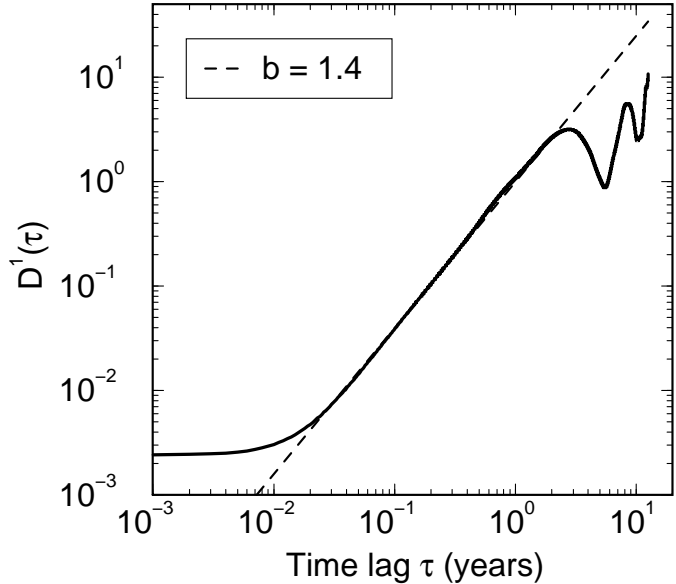}
\caption{\label{sf} Structure function of the simulated light curve
displayed in the middle panel of Figure~\ref{sim_lc}, corresponding to
the redshift and estimated Doppler factor of 3C~345. The fitted slope
of 1.4 (dashed line) provides a good match to that of the observed
structure function of 3C~345 at 4.8 GHz. }
\end{figure*}

A further comparison with the data can be made by examining the first
order structure function of our simulated light curves. This function
is defined as $D^1(\tau) = \left<[S(t) - S(t+\tau)]^2\right>$, where
$S(t)$ is the flux density at time $t$ and $\tau$ is the time
lag. Structure functions are commonly used as a means of
characterizing the distribution of power in a stochastic process on
different time scales (e.g., \citealt*{SCH85}). The slope $b =
d\log{D^1(\tau)} / d\log{\tau}$ is a general indicator of the type of
stochastic process, with $b=0$ and $b=1$ corresponding to flicker and
shot noise, respectively. As pointed out by
\cite{HAA92}, a slope of unity at long time lags is indicative of a
shot-noise process of the type we are simulating. These authors found
that the majority of the AGNs in the U. Michigan monitoring survey
have structure function slopes that cluster near unity, with values
ranging from approximately 0.7 to 1.65. In Figure~\ref{sf} we show the
structure function for the simulated light curve in the middle panel
of Figure~\ref{sim_lc} with $\delta = 27$ and $z = 0.595$,
corresponding to the properties of 3C~345. We find good agreement in
the structure function slope of our model ($b = 1.4$) and that
measured by \cite{HAA92} at 4.8 GHz.  The slope of our simulated
structure function for 3C~345 is slightly steeper than the value of
unity expected for a pure shot noise process, due to the Doppler
contraction of flaring timescales in the observer's frame, and the
fact that we are modeling only major flares, which have a relatively
low rate of occurence. The light curves of other AGNs may include
longer-term, low amplitude flares that serve to flatten the slope of the
structure function. We do not include these flares in our model, as
they simply add a slowly varying component to the quiescent level, and
do not have a major impact on the duty cycle of the source. 

\section{Variability duty cycles of beamed AGNs}
\begin{figure*}
\plotone{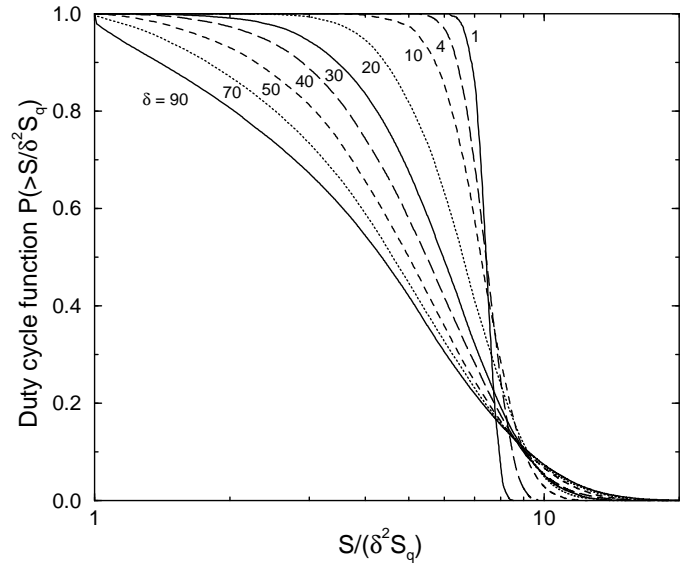}
\caption{\label{dutycycle1} Observed variability duty cycle functions of
an AGN at a redshift $z=1$ for different values of the jet Doppler
factor ($\delta$).  The curves from left to right represent values of
$\delta =$ 90, 70, 50, 40, 30, 20, 10, 4, and 1, respectively. Each
curve represents the fraction of time the source spends above a flux
density $S$, measured in units of the observed quiescent (non-flaring)
flux density ($\delta^2 S_q$). }
\end{figure*}

An important quantity in the statistics of flaring source
samples is the variability duty cycle, which we define to be the
fraction of time the source spends above a particular flux density
level.  In this section we derive numerical estimates of AGN duty
cycle functions by simulating light curves for sources with a wide
range of Doppler factors and redshifts. A list of our individual model
parameters is given in Table~\ref{parameters}. For each set of
parameters we simulate 500 twenty-year light curves, and generate a
cumulative histogram that gives the fraction of time spent above a
particular flux density $S$, measured in units of the observed
quiescent flux density (in our case, $\delta^2 S_q$). We plot the simulated
duty cycles for sources at $z=1$ in
Figure~\ref{dutycycle1}.  It is apparent that relativistic beaming can
have a large effect on the observed duty cycles of AGNs, especially
for sources with $\delta \gtrsim 10$. As expected, these highly beamed
jets tend to have a much wider range of observed variability amplitude
than the more weakly beamed sources.

\begin{figure*}
\plotone{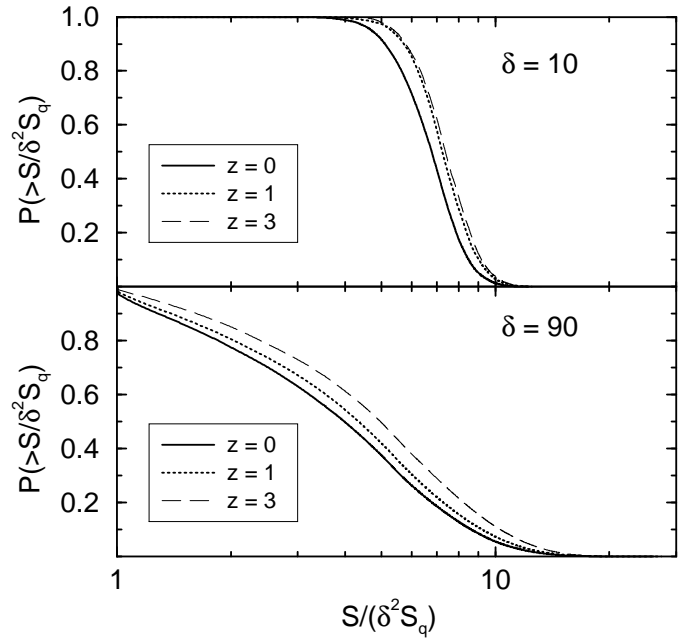}
\caption{\label{dutycycle3} Plots showing the influence of source
redshift on the duty cycle function of an AGN with Doppler factor 10
(top panel), and Doppler factor 90 (lower panel). The three curves in
each panel represent the observed duty cycle function for the source
at a redshift of $z = 0$, 1 and 2, respectively.  Sources at higher
redshifts experience greater time dilation in the observer frame,
which increases the probability of having flares that overlap in
time. This tends to raise the median steady-state flux level of the
source.  }
\end{figure*}

\begin{figure*}
\plotone{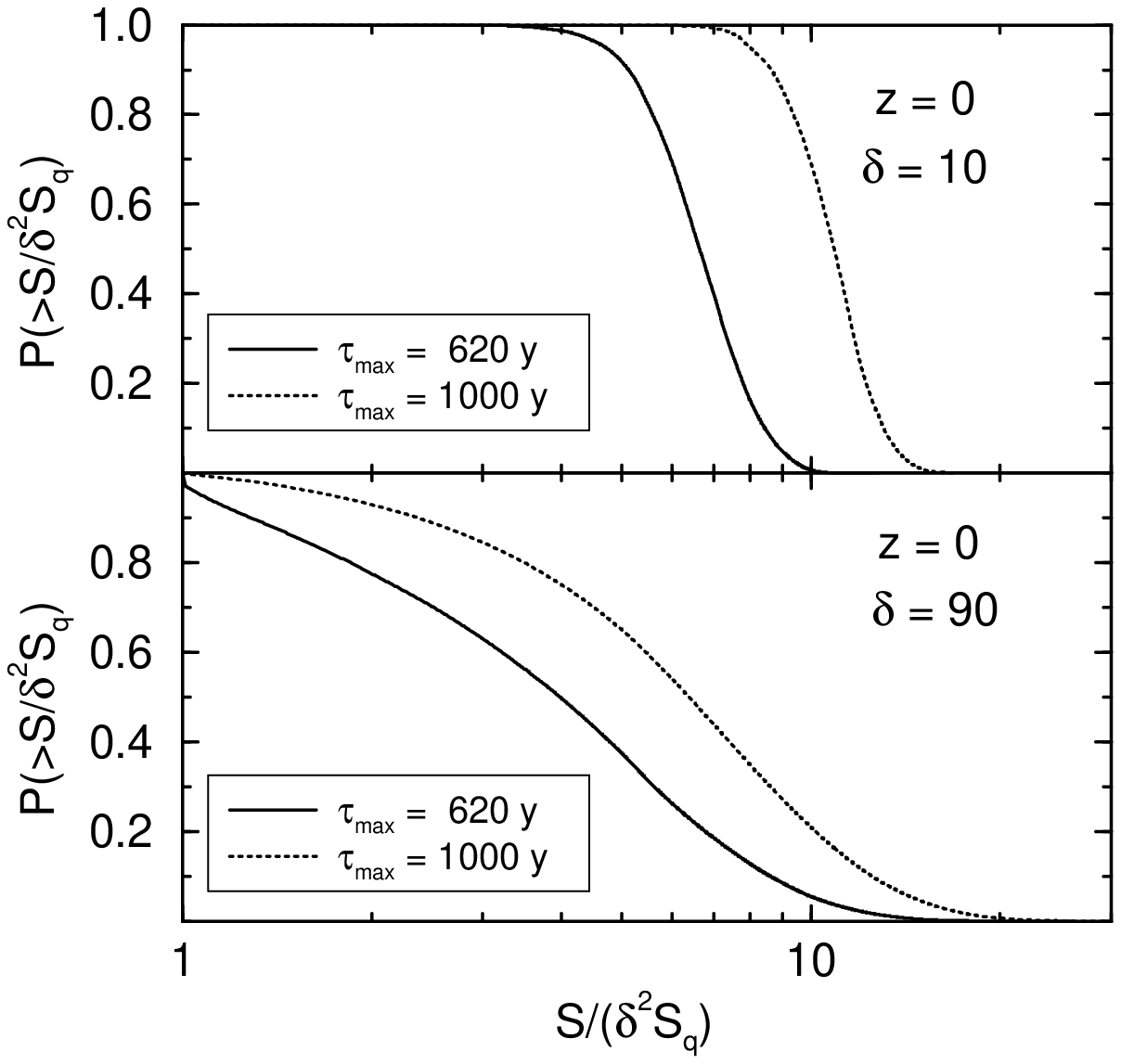}
\caption{\label{dutycycle2} Plots showing the influence of maximum
flare time scale on the duty cycle function of a source at redshift
zero with Doppler factor 10 (top panel), and Doppler factor 90 (lower
panel). Increasing the maximum time scale from 620 years to 1000 years
raises the probability of having flares that overlap in time, which
increases the median flux level of the source. }
\end{figure*}

The cosmological redshifts of AGNs will stretch their intrinsic flux
variations in time by a factor of $(1+z)$, and should therefore have
an impact on their observed duty cycles as well. In
Figure~\ref{dutycycle3} we show the duty cycle functions of two
simulated sources, one with $\delta = 10$ (upper panel), and the other
with $\delta = 90$ (lower panel). The three curves in each panel
represent the duty cycle of the source as it would appear at redshifts
of 0, 1, and 3. For sources located at high redshifts, cosmological
time dilation increases the probability that flares will overlap in
the observer frame. This tends to increase the median flux density
level of the source and shift the duty cycle curves to the right in
Figure~\ref{dutycycle3}.  A similar effect occurs when the maximum
flare time scale is increased (Figure~\ref{dutycycle2}), since this
will also increase the likelihood of overlapping flares. The net
effect of raising $\tau_{max}$ is to boost the overall flux density
levels of both strongly and weakly beamed sources alike. Although this
will have implications on the intrinsic luminosity function of the
parent population, it will not affect the statistics of flux-limited
samples.

Our duty cycle simulations reveal a somewhat paradoxical property of
AGN variability: although the flux densities of highly beamed sources
can reach very high levels due to boosting, their flaring timescales
are severely contracted, so that these sources are observed to spend
most of their time in relatively low flaring states. Consequently, it
is the {\it weakly beamed} sources that are more likely to be found in
a state that is much higher than their quiescent level, due to the
overlap of many long term flares.  As we will show in the following
section, this tends to slightly weaken the predicted effects of
Doppler bias in flaring samples selected on the basis of
relativistically beamed flux.

\section{\label{vareffects}Effects of variability on core-selected 
AGN samples}

The statistics of flux-limited samples should not be strongly influenced
by flux variability as long as the variability properties of the
parent population are relatively uniform. However, if some objects
have a markedly different variability duty cycle than the rest of the
population, this can affect the likelihood that they will appear in a
flux-limited sample, thereby creating a selection bias.

In this section we investigate possible variability biases in
flat-spectrum AGN samples by performing Monte Carlo simulations of
both flaring and non-flaring beamed jet populations.  We compare
the observed properties of flux-limited samples drawn from these
parent populations to determine to what degree core-selected AGN
samples are additionally biased by source variability.

\subsection{\label{nonflaring} Simulated non-flaring jet population}

In a previous study (\citealt{LM97, L99}), we successfully modeled the
observed radio properties of a complete core-selected AGN sample (the
Caltech-Jodrell Flat-spectrum survey; CJ-F) using a simulated
population of two-sided, relativistically beamed jets. We found that
the apparent jet velocity distribution of the CJ-F was best fit by a
parent population of jets with a power-law Lorentz factor distribution
that was weighted toward low speeds.  This also appears to be the case
for several other flat-spectrum AGN samples (e.g.,
\citealt{KVZ00,JMM01}) whose apparent velocity distributions are
similar to that of the CJ-F.

For the purposes of this study we adopt the best-fit model of
\cite{L99} for the CJ-F, which has no dependence of intrinsic
luminosity on jet speed and $N(\Gamma) \propto \Gamma^{-1.25}$, with
$1.001252 \le \Gamma \le 45$. The jets have random spatial
orientations and are distributed with a constant co-moving space
density out to $z = 4$. Their unbeamed 5 GHz luminosity function
follows a fit to that of powerful (FR-II) radio galaxies, which
incorporates pure exponential luminosity evolution \citep{UP95}.

Since few constraints exist on the size of the CJ-F parent population,
the lower cutoff of the parent luminosity function ($L_1$) is an
important parameter. We can obtain an estimate of $L_1$ by examining
the properties of the radio galaxy NGC 3894, which has the lowest 5
GHz luminosity of any source in the CJ-F sample. Its structure consists of a
two-sided jet on parsec-scales, which has been monitored extensively by
\cite{TWV98}. By using the apparent expansion and flux density ratio
of the jet and counter-jet, these authors found the viewing
angle and jet speed to be $\theta \simeq 50\arcdeg$ and $\beta
\simeq 0.3$, respectively. Assuming equal intrinsic luminosities and
spectral index of zero for the jet ($j$)  and counter-jet ($cj$), this implies that
the radio luminosity of NGC 3894 is boosted by a factor of $(\delta_j^p +
\delta_{cj}^p) \simeq 2 $, where we have  assumed $p = 2$, and
$\delta_j = 1.18$ and $\delta_{cj} = 0.8$.  Given that the
observed 5 GHz luminosity of NGC 3894 is $1.9 \times 10^{23} \;
\rm W\; Hz^{-1}$ \citep{TVR96}, we will adopt a lower luminosity
function cut-off of $9.5 \times 10^{22}\; \rm W\; Hz^{-1}$ for our
simulated non-flaring parent population.

In our Monte Carlo procedure we iteratively generate jets according to
the above parameters until we obtain a sample of 293 sources with
$S_{\rm5\; GHz} > 0.35 $ Jy.  These values correspond to the sample
size and flux density cutoff of the CJ-F. Approximately $2 \times
10^6$ parent objects are needed in a typical Monte Carlo run. In
\cite{LM97} we required a substantially larger parent population
($\sim 10^7$ objects) to fit the CJ-F, since we used a smaller value
of $L_1$ and a jet Lorentz factor distribution that only extended up
to $15 \; h^{-1}$.

\begin{figure*}
\plotone{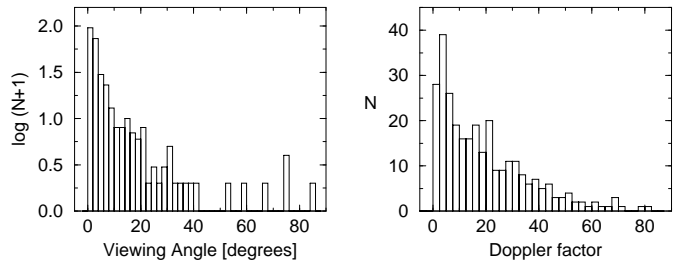}
\caption{\label{distribs} Predicted distributions of jet viewing angle (left
panel; log scale) and Doppler factor (right panel; linear scale) for
the Caltech-Jodrell Flat-spectrum survey (CJ-F) based on a non-variable
parent population. }
\end{figure*}

We show the viewing angle and Doppler factor distributions for a
single Monte Carlo run in Figure~\ref{distribs}. As expected, the
simulated CJ-F sample is heavily biased towards sources with low
viewing angles since these tend to have the highest amount of Doppler
boosting. The predicted Doppler factors of the CJ-F sources range up
to the maximum possible value of $\sim 2 \Gamma_{max} = 90$.

\subsection{Simulated flaring jet population}

In order to simulate a flaring jet population, only slight
modifications to our Monte Carlo model are necessary. We first assume
that all of the parent objects have the same intrinsic variability
characteristics given by the model parameters in
Table~\ref{parameters}. Although this is not likely true of the actual
AGN population, this simplification makes it easier to isolate the
effects of beaming on the statistics of flaring samples. Furthermore,
the fact that radio variability is well-correlated with several other
statistical beaming indicators (e.g., \citealt{AAH92, LTP01}) suggests
that the observed variability characteristics of AGNs are more
dependent on beaming than on intrinsic factors.

As in our previous Monte Carlo simulations, we simulate quiescent flux
levels for the parent objects, but we now also assign a flaring level
to each source based on its duty cycle function. We accomplish this by
first binning the source into one of four redshift bins at
$z$ = 0, 0.5, 1, 2, or 3, and then further binning it into one of
thirteen sub-bins based on its Doppler factor. We then randomly
generate a flare level according to the duty cycle (i.e., flaring
probability) curve for the appropriate redshift/Doppler factor
bin. The flux density of the source is then simply the quiescent
flux density times the flaring level. If this  flux density is
fainter than the CJ-F cutoff, the source is discarded. We continue
generating simulated sources in this manner until we obtain the
necessary 293 sources with $S_{\rm 5\; GHz} > 0.35 $ Jy.

We note that it is necessary in these flaring simulations to modify the
lower cutoff of the parent luminosity function, since we showed in
\S~4 that weakly beamed flaring sources are likely to be observed at
levels considerably higher than their quiescent (non-flaring)
level. Recalling that we used the least luminous source in the CJ-F
(NGC 3894) to estimate $L_1$ in the non-flaring case, we estimate from
our duty-cycle curves that its typical flaring level is approximately
7.4 times its quiescent level, given its redshift and Doppler factor.
We therefore decrease $L_1$ by this factor and adopt a lower cutoff of
$1.28 \times 10^{22}\; \rm W\; Hz^{-1}$ for the quiescent luminosities
of the simulated flaring population.

\subsection{Comparison of flaring and non-flaring samples}

\begin{figure*}
\plotone{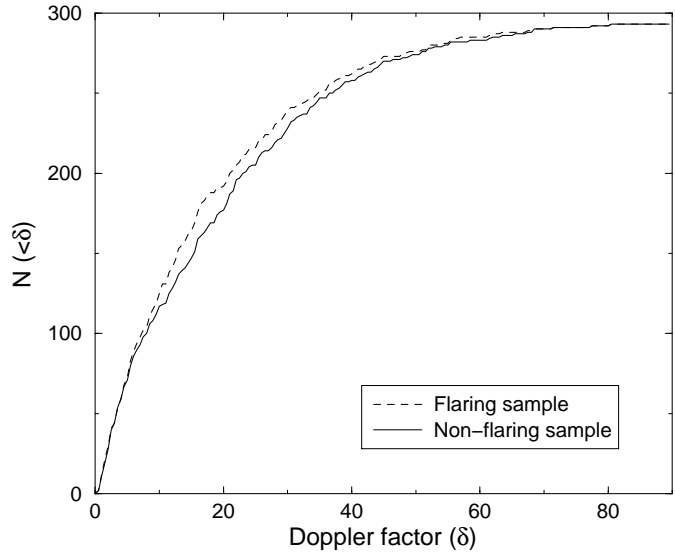}
\caption{\label{cumhist1} Predicted cumulative Doppler factor distributions  
for the CJ-F survey based on flaring (dashed line) and non-flaring
(solid line) models. The sample selected from a flaring population
is slightly more biased towards low-Doppler factor sources. }
\end{figure*}

In Figure~\ref{cumhist1} we show cumulative histograms of the
predicted Doppler factor distribution for the CJ-F sample in the case
of flaring (dashed line) and non-flaring (solid line) populations. The
differences in the two cases are rather small, with the flaring sample
being slightly more biased towards lower Doppler factors. There is
also slightly less orientation bias present in the flaring sample (not
shown), with relatively more sources found at higher viewing angles.
We have verified these results using multiple Monte Carlo runs that
all display the same effect. These findings are independent of our
flaring amplitude parameter $P$, since increasing this parameter
merely increases the overall flux density level of the entire parent
population. This will not affect the selection statistics since
increasing $P$ also requires that $L_1$ be adjusted downward
accordingly.

Although we have indeed found a bias in the statistics of flaring AGN
samples, its predicted effects are rather small in medium-sized
surveys such as the CJ-F.  The main reason for this lies with the
Lorentz factor distribution of the AGN population, which is weighted
toward slow jets. The majority of the CJ-F parent objects are
low-Doppler factor sources with jet axes inclined at $\sim 60\arcdeg$
from the line of sight. As we showed in \S~4, the duty-cycle functions
of these weakly-beamed objects are very similar, and are roughly
symmetric about the mean steady-state level (Fig.~\ref{dutycycle1}). A
sample composed predominantly of weakly-beamed flaring sources will
therefore be statistically similar to a non-flaring sample, since the
effects of variability will be averaged out.

Another important factor that reduces the size of the variability bias
is the limited range of flux levels seen in individual AGNs at radio
wavelengths (maximum peak/trough ratio $\sim 30$). By comparison, the
entire parent population likely spans a flux density range of at least
ten orders of magnitude, given the wide range of intrinsic
luminosities, cosmological distances, and beaming factors that are
present.  Variability will therefore only influence the selection
statistics of a relatively small fraction of the parent population
that have fluxes near the survey limit. In large core-selected samples
having a faint flux cutoff, these sources will tend to be weakly
beamed, due to the predicted decrease in mean Doppler factor with flux
density (see \citealt{L99}). As a result, the effects of variability
will be statistically averaged out as described above. Those sources
whose duty cycles are affected by beaming (with $\delta \gtrsim 10$)
will tend to have beamed flux densities that are well above the survey
limit, and will always be selected regardless of their flaring
level. We have verified these effects by performing simulations using
the same parent population as the CJ-F, only with a survey cutoff ten
times fainter. We find virtually no statistical differences in the
flaring and non-flaring samples, which contained approximately $\sim
3600$ sources.

\section{Conclusions}

We have developed a simple shot-noise variability model to examine how
relativistic beaming affects the radio light curves and selection
statistics of flat-spectrum AGNs.  We summarize our main findings as
follows:

1. Relativistic beaming will preferentially boost the amplitude of a
flaring event with respect to an AGN's quiescent flux level in the
observer frame, and will also shorten its apparent time scale. As a
result, highly beamed AGNs (i.e., with Doppler factors $\gtrsim
10$) are able to reach very high flux levels, but will in fact be
observed to spend most of their time in relatively low flaring states.

2. The fact that the most highly beamed blazars are not observed to
return to a quiescent level between flares suggests that the intrinsic
timescales of individual flares are rather long, given the large
Lorentz factors ($\Gamma \simeq 45$) of these jets that are inferred
from superluminal motion studies. We find that the flare amplitudes
must be a small fraction ($\lesssim 5\%$) of the quiescent flux level
in the source frame in order to reproduce the relatively small
peak-to-trough flux density ratios seen in the most highly variable
AGNs.  The unbeamed counterparts of blazars should therefore have very
stable radio light curves that are made up of a steady-state
superposition of many long-term flares. The apparent ``quiescent''
flux density levels of these sources (identified in many unified
models as radio galaxies) will be many times greater than their
``true'' quiescent (i.e., non-flaring) levels.

3. We have used the Monte Carlo beaming model of \cite{LM97} to model the
parent population of the Caltech-Jodrell Flat-spectrum survey (CJ-F)
using both flaring and non-flaring jets. We find that the standard
orientation bias toward highly beamed jets with small viewing angles
is predicted in both cases. In the case of a flaring population,
however, the amount of orientation bias is slightly reduced due to the
fact that the highly-beamed sources have a higher probability of being
found in low flaring states. The magnitude of this effect is rather
small in moderately-sized samples such as the CJ-F ($N \sim 300$
sources) due to the fact that the majority of the highly beamed
sources in the parent population lie well above the survey flux limit
and will be selected regardless of their flaring levels. We find that
for larger samples with fainter cutoffs, any selection biases
associated with variability should be negligible.

\acknowledgments

The author thanks A. P. Marscher and the referee, P. A. Hughes for
useful comments. This research has made use of data from the
University of Michigan Radio Astronomy Observatory, which is supported
by the National Science Foundation and by funds from the University of
Michigan.

\begin{deluxetable}{lll}
\tablecolumns{3}
\tablecaption{\label{parameters}AGN Variability Model Parameters}
\tablewidth{0pt}
\tablehead{\colhead{Symbol} & \colhead{Parameter} &\colhead{Value} }
\startdata
$\rho$       & Flaring rate &   2                    $\rm y^{-1}$ \\
$\tau_{min}$ & Minimum flare time scale &            1   month \\
$\tau_{max}$ & Maximum flare time scale &          620 y \\
$P$          & Peak-to-quiescent flux ratio of flare & 1.05  \\
$\delta$ & Doppler factor & 1, 2, 4, 6, 10, 20, 30, 40, 60, 70, 80, 90\\
$z$          & Redshift  & 0, 0.5, 1, 2, 3 \\
\enddata 
\tablecomments{The flaring rate, time scale, and peak-to-quiescent
ratio  are all defined in the AGN rest frame.}
 \end{deluxetable}


\begin{thebibliography}{}

\bibitem[Aller et al.(1992)Aller, Aller, \& Hughes(1992)]{AAH92} Aller, M.\ 
F., Aller, H.\ D.\ \& Hughes, P.\ A.\ 1992, \apj, 399, 16 

\bibitem[Aller et al.(1999)]{AAH99} Aller, M.\ F., Aller, H.\ D., Hughes, P.\ A., \& Latimer, G.\ E.\ 1999, \apj, 512, 601 

\bibitem[Blandford \& K\"onigl(1979)]{BK79} Blandford, R.\ D.\ 
\& K\"onigl, A.\ 1979, \apj, 232, 34 

\bibitem[Cruise \& Dodds(1985)]{CD85} Cruise, A.\ M.\ \& 
Dodds, P.\ M.\ 1985, \mnras, 215, 417 

\bibitem[Fujisawa et al.(1999)]{FKW99} Fujisawa, K., 
Kobayashi, H., Wajima, K., Hirabayashi, H., Kameno, S., \& Inoue, M.\ 1999, 
\pasj, 51, 537 

\bibitem[Hartman et al.(1999)]{H99} Hartman, R.\ C.\ et al.\ 1999,
 \apjs, 123, 79 

\bibitem[Hufnagel \& Bregman(1992)]{HB92} Hufnagel, B.\ R.\ 
\& Bregman, J.\ N.\ 1992, \apj, 386, 473 

\bibitem[Hughes et al.(1992)Hughes, Aller, \& Aller(1992)]{HAA92} Hughes, P.\ 
A., Aller, H.\ D., \& Aller, M.\ F.\ 1992, \apj, 396, 469

\bibitem[Jorstad et al.(2001a)]{JMM01} Jorstad, S. G., Marscher, A. P.,
Mattox, J. R., Wehrle, A. E., Bloom, S. D., Yurchenko, A. V. 2001a,
\apjs, 134,181

\bibitem[Jorstad et al. (2001b)]{JMMA01} Jorstad, S. G., Marscher,
A. P.,  Mattox, J. R., Aller, M. F., Aller, H. D., Wehrle, A. E., and
Bloom, S. D. 2001b, \apj, in press

\bibitem[Kellermann et al.(2000)]{KVZ00} Kellermann, K. I., Vermeulen,
R. C., Zensus, J. A., Cohen, M. H. 2000, in Astrophysical
Phenomena Revealed by Space VLBI, eds. H. Hirabayashi, P. G. Edwards,
\& D. W. Murphy (Sagamihara: Institute of Space \& Astronautical
Science), 159

\bibitem[L{\"a}hteenm{\"a}ki \& Valtaoja(1999)]{LV99} 
L{\"a}hteenm{\"a}ki, A.\ \& Valtaoja, E.\ 1999, \apj, 521, 493 

\bibitem[Lister(1999)]{L99} Lister, M. L. 1999, Ph.D. thesis, Boston
University. 

\bibitem[Lister \& Marscher(1997)]{LM97} Lister, M.\ L.\ \&  Marscher,
A.\ P.\ 1997, \apj, 476, 572

\bibitem[Lister et al.(2001)]{LTP01} Lister, M. L., Tingay, S. J., \&
Preston, R. A. 2001, \apj, 554, 964

\bibitem[Perlmutter et al.(1999)]{P99} Perlmutter, S.\ et 
al.\ 1999, \apj, 517, 565 

\bibitem[Polatidis \& Wilkinson(1998)]{PW98} Polatidis, 
A.\ G.\ \& Wilkinson, P.\ N.\ 1998, \mnras, 294, 327 

\bibitem[Scheuer \& Readhead(1979)]{SR79} Scheuer, P.\ A.\ 
G.\ \& Readhead, A.\ C.\ S.\ 1979, \nat, 277, 182 

\bibitem[Simonetti, Cordes, \& Heeschen(1985)]{SCH85} 
Simonetti, J.\ H., Cordes, J.\ M., \& Heeschen, D.\ S.\ 1985, \apj,
296, 46 

\bibitem[Taylor et al.(1996)]{TVR96} Taylor, G.\ B., 
Vermeulen, R.\ C., Readhead, A.\ C.\ S., Pearson, T.\ J., Henstock, D.\ R., 
\& Wilkinson, P.\ N.\ 1996, \apjs, 107, 37 

\bibitem[Taylor et al.(1998)Taylor, Wrobel, \& Vermeulen(1998)]{TWV98}
Taylor, G.\ B., Wrobel, J.\ M., \& Vermeulen, R.\ C.\ 1998, \apj, 498, 619 

\bibitem[Ter\"asranta \& Valtaoja(1994)]{TV94} Ter\"asranta, H.\ 
\& Valtaoja, E.\ 1994, \aap, 283, 51 

\bibitem[Urry \& Padovani(1995)]{UP95} Urry, C.\ M.\ \& 
Padovani, P.\ 1995, \pasp, 107, 803 

\bibitem[Valtaoja et al.(1999)]{VLT99} Valtaoja, E., L{\"a}hteenm{\"a}ki, A., 
Ter{\"a}sranta, H., \& Lainela, M.\ 1999, \apjs, 120, 95 

\bibitem[Valtaoja et al.(1992)]{VTUN92} Valtaoja, E., 
Ter\"asranta, H., Urpo, S., Nesterov, N.\ S., Lainela, M., \& Valtonen, M.\ 
1992, \aap, 254, 80 

\bibitem[Vermeulen \& Cohen(1994)]{VC94} Vermeulen, R.\ C.\ 
\& Cohen, M.\ H.\ 1994, \apj, 430, 467 

\bibitem[Wardle et al.(1994)]{WCR94} 
Wardle, J.\ F.\ C., Cawthorne, T.\ V., Roberts, D.\ H., \& Brown, L.\ F.\ 
1994, \apj, 437, 122 

\end{thebibliography}
\end{document}